\documentclass[journal,10pt]{IEEEtran} 
\usepackage[figurename=Fig.]{caption}
\usepackage{tikz}
\usetikzlibrary{calc, positioning, shapes.geometric, decorations.pathreplacing}
\usepackage{array}
\usepackage{verbatim}
\usetikzlibrary{calc, shapes, arrows, positioning}
\usepackage{authblk}
\usepackage{blindtext}
\usepackage{ifpdf}
\usepackage{graphicx}
\usepackage{tabulary}
\usepackage{amsmath,amsthm,amssymb}
\usepackage{amsmath}
\usepackage{kantlipsum}
\allowdisplaybreaks
\usepackage{cleveref}
\usepackage{lipsum}%
\usepackage{optidef}
\usepackage[english]{babel} 
\theoremstyle{plain}
\newtheorem{theorem}{Theorem}

\crefname{theorem}{Theorem}{theorem}
\crefname{lemma}{Lemma}{Lemmas}

\usepackage{cite}
\usepackage{dsfont}
\usepackage[justification=centering]{caption}
\usepackage{color}
\usepackage{algorithm}
\usepackage{algorithmicx, algpseudocode}
\usepackage{etoolbox}
\usepackage{subcaption}
\usepackage{balance}
\usepackage{empheq,etoolbox}
\usepackage[utf8]{inputenc}
\usepackage{multirow}
\usepackage{float}

\usepackage{amssymb}
\usepackage{pifont}
\usepackage[hmargin=1.27cm,top=1.3cm,bottom=1.3cm]{geometry}

\usepackage{physics}

\usepackage{babel}
\floatstyle{plaintop}
\restylefloat{table}
\patchcmd{\subequations}
\usepackage{lipsum} 
\allowdisplaybreaks

\tikzset{brace/.style={decorate, decoration={brace}},
 brace mirrored/.style={decorate, decoration={brace,mirror}},
}

\newcounter{brace}
\newcounter{arrow}
\begin{document}
 \captionsetup[figure]{name={Fig.},labelsep=period}

\title{Energy-Efficient Resource Allocation for NOMA-Assisted Uplink Pinching-Antenna Systems}

\bstctlcite{IEEEexample:BSTcontrol}
\author{Ming Zeng, Xingwang Li, Ji Wang, Gaojian Huang, Octavia A. Dobre, \textit{Fellow, IEEE} and Zhiguo Ding, \textit{Fellow, IEEE}
    \thanks{M. Zeng is with Laval University, Quebec City, Canada (email: ming.zeng@gel.ulaval.ca).}

    \thanks{X. Li is with Henan Polytechnic University, Jiaozuo, China (email: lixingwang@hpu.edu.cn).}
    
    \thanks{J. Wang is with Central China Normal University, Wuhan, China (e-mail: jiwang@ccnu.edu.cn).}

    \thanks{G. Huang is with Henan Polytechnic University, Jiaozuo, China (e-mail: g.huang@hpu.edu.cn).}


\thanks{O. A. Dobre is with the Faculty of Engineering and Applied Science, Memorial University, St. John’s, Canada (e-mail: odobre@mun.ca).}

\thanks{Z. Ding is with Khalifa University, Abu Dhabi, UAE, and the University of Manchester, Manchester, UK. (e-mail:zhiguo.ding@ieee.org).}   
    }
\maketitle

\begin{abstract}
The pinching-antenna architecture has emerged as a promising solution for reconfiguring wireless propagation environments and enhancing system performance. While prior research has primarily focused on sum-rate maximization or transmit power minimization of pinching-antenna systems, the critical aspect of energy efficiency (EE) has received limited attention. Given the increasing importance of EE in future wireless communication networks, this work investigates EE optimization in a non-orthogonal multiple access (NOMA)-assisted multi-user pinching-antenna uplink system. The problem entails the joint optimization of the users' transmit power and the pinching-antenna position. The resulting optimization problem is non-convex due to tightly coupled variables. To tackle this, we employ an alternating optimization framework to decompose the original problem into two subproblems: one focusing on power allocation and the other on antenna positioning. A low-complexity optimal solution is derived for the power allocation subproblem, while the pinching-antenna positioning subproblem is addressed using a particle swarm optimization algorithm to obtain a high-quality near-optimal solution. Simulation results demonstrate that the proposed scheme significantly outperforms both conventional-antenna configurations and orthogonal multiple access-based pinching-antenna systems in terms of EE.

\end{abstract}

\IEEEpeerreviewmaketitle

\section{Introduction}
Owing to the abundant spectrum availability at high frequencies, bands such as millimeter-wave and terahertz are critical for addressing the ever-growing capacity demands of next-generation wireless communication systems \cite{Sun_TVT18, Hao_Network22}. However, these high-frequency channels are characterized by severe attenuation and poor diffraction capability, making them predominantly line-of-sight (LoS) dependent. Consequently, the presence of an unobstructed LoS path is crucial for reliable performance in high-frequency bands systems. In practical deployments, LoS links may be subject to obstacles, leading to potential system outages. To mitigate this issue, NTT DOCOMO introduced the concept of pinching-antenna systems in 2022 \cite{Atsushi_22}, wherein an antenna structure was formed by applying a plastic pinch to a dielectric waveguide. On this basis, a strong LoS path can be established, enabling robust video transmission in the high-frequency band.

Since the initial demonstration by NTT DOCOMO, pinching-antenna systems have emerged as a promising solution for establishing LoS connectivity in scenarios where users would experience non-line-of-sight (NLoS) channel conditions without pinching antennas \cite{ding2024, yang2025, liu2025pinching}. To fully harness the benefits of pinching antennas, their spatial placement must be jointly optimized with conventional wireless resources such as time, frequency, and power \cite{tyrovolas2025, Xu_WCL25}. In \cite{tyrovolas2025}, the authors investigated a single-user system assisted by a single pinching antenna. The outage probability and the average achievable data rate were analyzed, and the antenna placement was subsequently optimized to enhance system performance. 
Building upon this, \cite{Xu_WCL25} extended the framework to a multi-antenna setup, where multiple pinching antennas collaboratively serve a single user. Their results demonstrated that optimal placement involves minimizing path loss while ensuring constructive signal superposition to maximize the user's transmission rate.

In multi-user scenarios, the integration of non-orthogonal multiple access (NOMA) into pinching-antenna systems is naturally facilitated, since the signal transmitted through a dielectric waveguide naturally consists of a superposition of signals from all the users being served \cite{wang2024, xie2025, Zeng_COMML25, fu2025}. Within this context, prior studies have primarily focused on maximizing the sum rate \cite{ wang2024, xie2025, Zeng_COMML25} or minimizing transmit power \cite{fu2025} in NOMA-enabled pinching-antenna systems.

However, as far as we are aware, the problem of energy efficiency (EE) maximization remains unexplored in the existing literature. As EE is expected to be a fundamental performance metric in next-generation wireless networks, it is imperative to investigate its optimization in the context of pinching-antenna architectures. In this work, we consider EE maximization in a NOMA-assisted multi-user uplink system incorporating a single antenna. The problem involves the joint optimization of the antenna’s spatial position and the users’ transmit power, resulting in a non-convex formulation with interdependent variables. To address this, we adopt an alternating optimization (AO) approach that decomposes the problem into two manageable subproblems: power allocation and antenna positioning. For the former, we derive a low-complexity optimal solution, while the latter is efficiently solved using the particle swarm optimization (PSO) algorithm to obtain a high-quality near-optimal placement. Simulation results verify that the proposed scheme is near-optimal and significantly enhances EE compared to both fixed-antenna baselines and orthogonal multiple access (OMA)-based pinching-antenna systems.

\begin{figure}[ht!]
\centering
\includegraphics[width=0.8\linewidth]{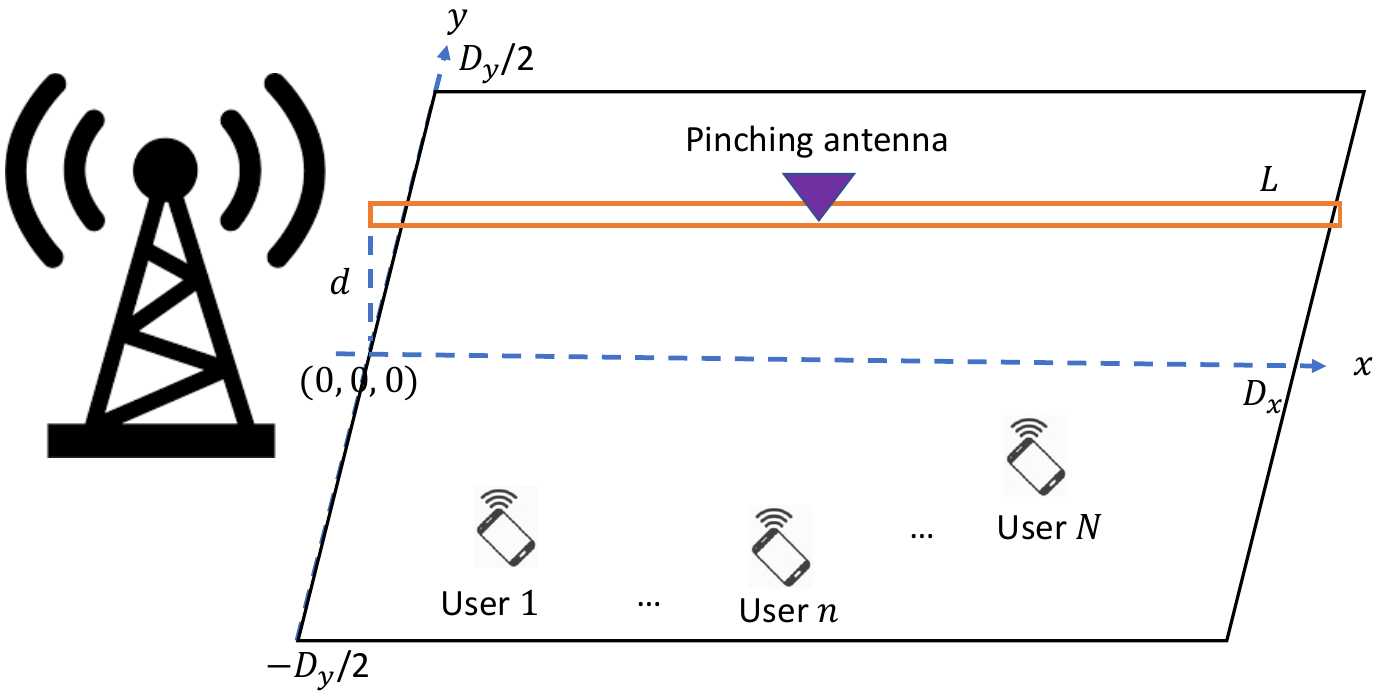}
\caption{Considered pinching-antenna system.} 
\label{fig:Low_2}
\end{figure}

\section{System Model and Problem Formulation}
\subsection{System Model}
As shown in Fig. 1, we consider an uplink communication scenario in which $N$ single-antenna users transmit data to an access point (AP) equipped with a pinching antenna. The pinching antenna is applied to a dielectric waveguide of fixed height $d$ and length $L$, and can be dynamically activated at any point along the waveguide. The users are assumed to be randomly located within a rectangular service area in the $x-y$ plane, with dimensions $D_x$ and $D_y$.  Let $\Psi^{\text{Ant}}=(x^{\text{Ant}}, 0, d)$ denote the 3D coordinate of the pinching antenna, and $\Psi_n= (x_n, y_n, 0)$ represents the position of the $n$th user, where $x_n \in [0, D_x]$ and $y_n \in [-D_y/2, D_y/2], \forall n$.


The users' channels are modeled using the free-space path loss model \cite{ding2024}. Accordingly, the channel gain for the 
$n$th user is given by
\begin{equation}
    h_n=\frac{c^2}{ 16 \pi^2 f_c^2  \abs{\Psi_i - \Psi^{\text{Ant}}}^2 }, 
\end{equation}
where $f_c$ and $c$ represent the carrier frequency and the speed of light, respectively. 

To serve multiple users simultaneously, we adopt NOMA, with successive interference cancellation performed at the AP to suppress inter-user interference. In typical uplink NOMA systems, users with stronger channel gains are decoded first. For notational simplicity, we assume that, for any given pinching-antenna position, users are indexed in a descending order of their channel gains, i.e., $h_1 \geq h_2 \geq \cdot \cdot \cdot \geq h_N$.\footnote{Actually, the SIC decoding order has no impact on the sum rate, as will be explained in the next section.} 
Based on this ordering, the achievable rate for the $n$th user is expressed as
\begin{equation}
    R_n^{\text{NOMA}}=\log_2 \bigg(1+  \frac{ P_n h_n }{ \sum_{i=n+1} ^N   P_i h_i  +  \sigma^2 } \bigg), 
\end{equation}
where $P_n$ denotes the transmit power of the $n$th user, subject to $P_n \leq P_n^{\max}$, with $P_n^{\max}$ being the maximum allowable power and $\sigma^2$ denotes the noise power at the AP. 

\subsection{Problem Formulation}
As demonstrated in \cite{Zeng_COMML25}, when the system objective is to maximize the sum rate, each user should transmit at full power. However, in practical scenarios where user terminals are subject to power constraints, optimizing EE offers a more balanced trade-off between system throughput and power consumption. Following the definition in \cite{Zhang_TVT17, Zeng_Access18, Zeng_COMML21}, EE is defined as the ratio of the system's sum rate to its total power consumption, which comprises both fixed circuit power and dynamic transmit power. Let $P_f$ denote the aggregate fixed circuit power of the system. Then, the EE can be expressed as \cite{Zhang_TVT17, Zeng_Access18, Zeng_COMML21} 
\begin{subequations}
\begin{align}
    \eta_{\text{EE}}&=\frac{\sum_{n=1}^N R_n^{\text{NOMA}}}{P_f+\sum_{n=1}^N P_n} \\
    &=\frac{\sum_{n=1}^N \log_2 \bigg(1+  \frac{ P_n h_n }{ \sum_{i=n+1} ^N   P_i h_i  +  \sigma^2 } \bigg)}{P_f+\sum_{n=1}^N P_n}.
\end{align}
\end{subequations}

Our objective is to maximize the EE by jointly optimizing the users’ transmit power and the position of the pinching antenna. This optimization problem can be formulated as:
\begin{subequations} \label{P_NOMA}
   \begin{align}
    \max_{x^{\text{Ant}}, P_n}~ & \eta_{\text{EE}} \\
    \text{s.t.}~& x^{\text{Ant}} \in [0, L], \\
    & P_n \leq P_n^{\max}, \forall n, 
\end{align} 
\end{subequations}
where constraint (\ref{P_NOMA}b) ensures that the pinching antenna remains within the bounds of the dielectric waveguide, while constraint (\ref{P_NOMA}c) restricts each user’s transmit power to not exceed its maximum allowable value.

\section{Proposed Solution}
It is evident that problem \eqref{P_NOMA} is non-convex due to the non-convex nature of the objective function in (\ref{P_NOMA}a). Furthermore, the coupling between the users' transmit power and the pinching antenna location adds additional complexity, making the problem challenging to solve directly. To address this, we employ an AO framework that decomposes the original problem into two subproblems: one focused on power allocation and the other on pinching antenna positioning. In the following sections, we analyze each subproblem in detail.

\subsection{Transmit Power Allocation at the Users}
Under given position of the pinching antenna, the power allocation subproblem can be expressed as
\begin{subequations} \label{P_NOMA_0}
   \begin{align}
    \max_{P_n}~ & \frac{\sum_{n=1}^N \log_2 \bigg(1+  \frac{ P_n h_n }{ \sum_{i=n+1} ^N   P_i h_i  +  \sigma^2 } \bigg)}{P_f+\sum_{n=1}^N P_n} \\
    \text{s.t.}~ & P_n \leq P_n^{\max}, \forall n.
\end{align} 
\end{subequations}

In its current form, the numerator of the objective function---the sum rate---is non-convex, which poses significant challenges for optimization. To facilitate a tractable analysis, we begin by reformulating the users' sum rate as follows \cite{Zeng_COMML21, Zeng_COMML25}:
\begin{subequations} \label{P_NOMA_PA}
   \begin{align}
    \sum_{n=1}^N R_n^{\text{NOMA}}&= \sum_{n=1}^N \log_2 \bigg(1+  \frac{ P_n h_n }{ \sum_{i=n+1} ^N   P_i h_i  +  \sigma^2 } \bigg) \\
    &= \log_2 \big(\sum_{n=1} ^N   P_n h_n  +  \sigma^2  \big),
\end{align} 
\end{subequations}
where all other terms are canceled out. Clearly, the user ordering has no impact on the sum rate, i.e., (\ref{P_NOMA_PA}b). 
Based on this simplification, the EE can be reformulated as
\begin{equation} \label{Sum_EE}
    \eta_{\text{EE}}= \frac{\log_2 \bigg(\sum_{n=1} ^N   P_n h_n  +  \sigma^2  \bigg) }{P_f+\sum_{n=1}^N P_n}.
\end{equation}

\begin{theorem} \label{theorem 1}
    To maximize the EE, a user should not transmit if there exists another user with a better channel condition whose optimal transmit power is strictly less than its maximum allowable power.
\end{theorem}
\begin{IEEEproof}
    We prove this by contradiction. Consider user $n$ and suppose there exists another user $i$ such that $h_n < h_i$, i.e., user $i$ has a stronger channel. Furthermore, assume that the optimal transmit power for user $i$ satisfies $P_i^{\text{Opt}}<P_i^{\max}$, and user $n$'s optimal power satisfies $P_n^{\text{Opt}}>0$. Now, consider decreasing the transmit power of user $n$ by an amount $\Delta P$, where $\Delta P \leq \min( P_n^{\text{Opt}}, P_i^{\max}- P_i^{\text{Opt}})$, and simultaneously increasing the transmit power of user $i$ by $\Delta P$. Since $h_i > h_n$, this power reallocation results in an increased sum rate while keeping the total transmit power---and hence total power consumption---unchanged. Consequently, the overall EE increases, which contradicts the assumption that the original power allocation was optimal. Therefore, the initial assumption must be false, and the optimal transmit power for user $n$ must be zero. This concludes the proof.
\end{IEEEproof}

Building upon Theorem 1, we propose a low-complexity algorithm to determine the optimal power allocation. The key idea is to allocate power to the users sequentially, in a descending order of their channel gains. We begin by considering user 1, the user with the strongest channel gain. According to Theorem 1, all other users are assigned zero power, i.e., $P_n=0, \forall n>1$. Next, for user 1, the EE maximization problem is formulated as follows:

\begin{subequations} \label{P_NOMA_PA_U1}
   \begin{align}
    \max_{P_1}~ & \frac{ \log_2 (P_1 h_1  +  \sigma^2  )}{P_f+P_1} \\
    \text{s.t.}~ & P_1 \leq P_1^{\max}. 
\end{align} 
\end{subequations}

Problem \eqref{P_NOMA_PA_U1} can be classified as a fractional programming problem, and hence can be efficiently solved using the Dinkelbach algorithm \cite{Dinkelbach}. Specifically, the original problem is transformed into a sequence of parametric subtractive-form subproblems, which can be expressed as follows:
\begin{subequations} \label{P_NOMA_PA_U1_2}
   \begin{align}
    \max_{P_1}~ &  \log_2 (P_1 h_1  +  \sigma^2  )-\beta^{(l-1)} ( {P_1+P_f} )\\
    \text{s.t.}~ & P_1 \leq P_1^{\max}, 
\end{align} 
\end{subequations}
where $\beta^{(l-1)}$ is a non-negative parameter. Starting from $\beta^{(0)}=\frac{\log_2( P_1^{\max} h_1 +  \sigma^2  ) }{P_1^{\max} +P_f }$, $\beta^{(l)}$ is updated by 
\begin{equation}
    \beta^{(l)}=\frac{ \log_2 (P_1^{(l)} h_1  +  \sigma^2  )}{P_1^{(l)}+P_f}, 
\end{equation}
where $P_1^{(l)}$ represents the updated power after solving problem \eqref{P_NOMA_PA_U1_2}. As demonstrated in \cite{Dinkelbach}, the parameter $\beta^{(l)}$ increases with each iteration, $l$. The iterations terminate when the difference $\beta^{(l)}-\beta^{(l-1)}$ becomes smaller than a predefined threshold, such as $10^{-6}$. At this point, the final value of $\beta^{(l)}$ corresponds to the maximum EE of problem \eqref{P_NOMA_PA_U1}.

The remaining task is to solve problem \eqref{P_NOMA_PA_U1_2}. Since the objective function is concave, the optimal solution is achieved either at the stationary point---where the first derivative equals zero---or at one of the boundary points. The transmit power that nulls the derivative is given by 
\begin{equation}
    P_1=\frac{1}{\beta^{(l-1)} \ln{2}}- \frac{ \sigma^2}{h_1}.
\end{equation}

If this value lies within the feasible interval 
$[0, P_1^{\max}]$, it is the optimal transmit power for the 
$l$th iteration. Otherwise, the optimal solution lies at the boundary: if $P_1 <0$, then $P_1 =0$ and if $P_1>P_1^{\max}$, then $P_1 =P_1^{\max}$. Thus, a closed-form solution is obtained for problem \eqref{P_NOMA_PA_U1_2}.


After determining the optimal transmit power, i.e., $P_1^{\text{Opt}} $ for user $1$, we proceed to user 2. According to Theorem 1, if $P_1^{\text{Opt}} < P_1^{\max}$, then the optimal transmit power for user $2$ and for all remaining users must be zero, and the algorithm terminates. Conversely, if $P_1^{\text{Opt}} = P_1^{\max}$, the optimal power for user $2$ can be computed using the same procedure as for user $1$. Specifically, the EE maximization problem for user $2$ can be formulated as: 
\begin{subequations} \label{P_NOMA_PA_U2}
   \begin{align}
    \max_{P_2}~ & \frac{ \log_2 (P_1^{\max} h_1+ P_2 h_2  +  \sigma^2  )}{P_f+P_1^{\max}+P_2} \\
    \text{s.t.}~ & P_2 \leq P_2^{\max}. 
\end{align} 
\end{subequations}

Similarly, problem \eqref{P_NOMA_PA_U2} can be transformed into a sequence of parametric subtractive-form subproblems. Let
\begin{equation}
    P_2^{\text{Der}}=\frac{1}{\beta^{(l-1)} \ln{2}}- \frac{P_1^{\max} h_1+ \sigma^2}{h_2}
\end{equation}
denote the transmit power that nullifies the derivative of the objective function in the $l$th subtractive-form subproblem. The optimal power allocation for user 2 in the $l$th iteration is then given by:
\begin{equation}
    P_2^{\text{Opt}}=
    \begin{cases}
    P_2^{\text{Der}}, \text{if}~ P_2^{\text{Der}} \in [0, P_2^{\max}]\\
    0, \text{if}~ P_2^{\text{Der}} <0\\
    P_2^{\max}, \text{if}~ P_2^{\text{Der}} >P_2^{\max}. 
    \end{cases}
\end{equation}

Likewise, if the optimal power of user 2, i.e., $P_2^{\text{Opt}}$, satisfies $P_2^{\text{Opt}} < P_2^{\max}$, the algorithm terminates, and the optimal transmit powers for all users 
$n, n>2$ are set to zero. Otherwise, the same procedure is recursively applied to determine the optimal power allocation for the remaining users. The complete procedure for EE maximization via optimal power allocation is summarized in Algorithm~\ref{alg:cap}.


\begin{algorithm} 
\caption{Proposed Optimal Power Allocation for EE maximization under Given Pinching-Antenna Location}\label{alg:cap}
\begin{algorithmic} 
\State \bf{Initialize} $\xi^{\star} \gets 10^{-6}$; 
\For{$n \gets 1$ to $N$}    
    \State $\xi \gets 1$; $\beta^{(0)} \gets \frac{\log_2( \sum_{i=1}^{n}P_i^{\max} h_i +  \sigma^2  ) }{\sum_{i=1}^{n}P_i^{\max} +P_f }$; $l \gets 1$
    \While{$\xi \geq \xi^{\star}$}
        \State $P_n^{{\text{Der}}} \gets \frac{1}{\beta^{(l-1)} \ln{2}}- \frac{ \sum_{i=1}^{n-1} P_i^{\max} h_i+ \sigma^2}{h_n}$;
        \State $P_n^{\text{Opt}} \gets
        \begin{cases} 
        P_n^{\text{Der}}, \text{if}~ P_n^{\text{Der}} \in [0, P_n^{\max}]\\
        0, \text{if}~ P_n^{\text{Der}} <0\\
        P_n^{\max}, \text{if}~ P_n^{\text{Der}} >P_n^{\max}. 
        \end{cases}$;
        \State $\beta^{(l)}=\frac{\log_2( \sum_{i=1}^{n-1}P_i^{\max} h_i + P_n^{\text{Opt}} h_n + \sigma^2  ) }{\sum_{i=1}^{n-1}P_i^{\max}+ P_n^{\text{Opt}} +P_f }$;
        \State $\xi \gets \beta^{(l)}-\beta^{(l-1)}$;
        \State $l \gets l+1$;
    \EndWhile
    \If{$P_n^{\text{Opt}} < P_n^{\max}$} 
        \State break; \Comment{terminate the for loop in advance}
    \EndIf
\EndFor
\end{algorithmic}
\end{algorithm}

\subsection{Optimization of Pinching-Antenna Location}
Given a fixed power allocation, the total power consumption becomes constant. As a result, maximizing EE reduces to maximizing the system sum rate. Based on \eqref{P_NOMA_PA}, the problem for determining the optimal pinching-antenna location to maximize the EE (or equivalently, the sum rate) can be formulated as:
\begin{subequations} \label{Sum_rate}
   \begin{align}
    \max_{x^{\text{Ant}}}~ &  \log_2 \bigg( \sum_{n=1}^N  P_n h_n   +  \sigma^2  \bigg)  \\
    \text{s.t.}~& x^{\text{Ant}} \in [0, L].   
\end{align} 
\end{subequations}

Expanding $h_n$ using its definition, we obtain
\begin{subequations}
\begin{align}
    h_n&=\frac{c^2/16 \pi^2 f_c^2 }{   \abs{\Psi_i - \Psi^{\text{Ant}}}^2 } \\
    &=\frac{c^2/16 \pi^2 f_c^2 }{  \abs{x^{\text{Ant}} -x_n }^2   + y_n^2+ d^2 }, 
\end{align}
\end{subequations}
where $x_n$ and $y_n$ denote user 
$n$'s coordinates along the $x$- and $y$-axes, respectively.

Substituting $h_n$ with the above expanded expression, the objective function in \eqref{Sum_rate} can be expressed as 
\begin{equation}
  \sum_{n=1}^N R_n^{\text{NOMA}}   =\log_2 \bigg( \frac{c^2}{16 \pi^2 f_c^2}  \times  \sum_{n=1}^N  \frac{P_n }{ \abs{x^{\text{Ant}} -x_n }^2   + y_n^2+ d^2 }   +  \sigma^2  \bigg). 
\end{equation}

Given that the logarithmic function is monotonically increasing, it follows that maximizing the sum rate is equivalent to maximizing
\begin{equation}
    \sum_{n=1} ^N   \frac{P_n} {\abs{x^{\text{Ant}} -x_n }^2   + y_n^2+ d^2}.
\end{equation}

Accordingly, the optimal location of the pinching antenna can be determined by solving the following optimization problem:
\begin{subequations} \label{Sum_Rate_3}
   \begin{align}
    \max_{x^{\text{Ant}} }~ &   \sum_{n=1} ^N  \frac{P_n} {\abs{x^{\text{Ant}} -x_n }^2   + y_n^2+ d^2}    \\
    \text{s.t.}~& x^{\text{Ant}} \in [0, L].
\end{align} 
\end{subequations}

Each term (\ref{Sum_Rate_3}a) represents a generalized bell-shaped membership function, characterized by a symmetric bell curve and inherently non-convex in nature \cite{Anuar_ISCAS19}. As demonstrated in \cite{Zeng_COMML25}, the sum of such functions can lead to multiple local optima. An effective approach for addressing this non-convex optimization problem is the PSO algorithm. Notably, \cite{Zeng_COMML25} shows that the solution obtained via PSO closely matches that of an exhaustive search, thereby validating its effectiveness. In this work, we adopt the PSO algorithm to optimize the pinching-antenna location. Due to space constraints and to avoid redundancy, the implementation details are omitted; interested readers are referred to Section III of \cite{Zeng_COMML25} for a comprehensive description.


\subsection{Alternating Optimization until Convergence}
The transmit power allocation subproblem \eqref{P_NOMA_0} and the pinching-antenna position optimization subproblem \eqref{Sum_rate} are solved alternately until convergence. For the transmit power allocation subproblem, the optimal solution is obtained analytically, ensuring that the EE either increases or remains unchanged after each power update. However, for the pinching-antenna position optimization subproblem, since the PSO algorithm provides a heuristic solution, there is no guarantee that the EE will continue to improve or remain constant after each antenna position update. To ensure convergence, if the updated pinching-antenna position from the PSO algorithm does not result in a higher EE, the previous position is retained, and the procedure is terminated. Subsequently, the EE value from the most recent transmit power allocation subproblem is adopted as the final solution.


\subsection{Initialization}
The joint optimization of the users' transmit powers and pinching-antenna location is a non-convex problem, meaning that both its convergence rate and the final solution are influenced by the initialization. Based on Theorem 1, users with better channel conditions should be prioritized for higher transmit power. Consequently, for initialization, we propose placing the pinching antenna at a location that minimizes its distance to the nearest user. To determine this optimal position, for a given user $n, \forall n$, we first compute $x_n^{\text{Ant}}=\min(L, x_n)$ by solving the following optimization problem
\begin{subequations} \label{Initialization}
   \begin{align}
    \min_{x^{\text{Ant}}}~ &\abs{x^{\text{Ant}} -x_n }^2   + y_n^2 \\
    \text{s.t.}~& x^{\text{Ant}} \in [0, L].
\end{align} 
\end{subequations}

Among the $N$ potential locations, we select the one that minimizes $\abs{x_n^{\text{Ant}} -x_n }^2   + y_n^2$. 




\section{Simulation Results}
In this section, simulations are carried out to assess the performance of the proposed NOMA-assisted pinching-antenna system. To highlight the advantages of the pinching-antenna design, a baseline fixed-antenna configuration---referred to as ``NOMA-Fixed''---is considered, where the antenna is fixed at the location 
$(0, 0, d)$ meters. To demonstrate the necessity of employing NOMA, a time division multiple access (TDMA) benchmarking scheme is also included, wherein each user is assigned an equal time slot (denoted as ``TDMA''). Furthermore, to validate the effectiveness of the proposed AO framework and initialization strategy, comparisons are made with two alternatives: (i) an optimal benchmark solution obtained via an exhaustive search over the antenna's possible locations (``NOMA-Exhaustive''), and (ii) an AO scheme initialized with a randomly selected pinching-antenna position (``NOMA-Random'').


The simulations adopt the following default parameters, as in \cite{xie2025, Zeng_COMML25}: an antenna elevation of $d=3$ m, a fixed power consumption of $P_f=10$ dBm, a carrier frequency of $f_c=28$ GHz as well as a noise power of $\sigma^2=-90$ dBm. A total of 5 users are uniformly and randomly distributed over a service area measuring $D_x=120$ m and $D_y=20$ m, subject to $P_n^{\max}=10$ dBm, $\forall n$. Furthermore, the dielectric waveguide length is configured to match the horizontal span of the service region, i.e., $L=D_x$. Simulation results are averaged over $10^3$ independent Monte Carlo trials to ensure statistical robustness. 

Figure 2 illustrates the EE as a function of the maximum transmit power constraint at the users. As anticipated, the EE initially increases and eventually saturates across all evaluated schemes. This behavior stems from the fractional formulation of EE and the diminishing marginal rate gains associated with the logarithmic rate-power relationship. 
Among the considered methods, the proposed approach closely tracks the performance of the exhaustive search benchmark and consistently outperforms the variant employing random pinching-antenna placement. A significant performance gap is observed between NOMA and TDMA, which can be attributed to NOMA’s ability to prioritize users with stronger channel conditions, whereas TDMA allocates equal time to each user regardless of channel quality. Finally, the NOMA scheme with a fixed antenna exhibits the lowest EE, underscoring the performance advantages of the dynamic pinching-antenna design.

{\color{black}
Figure 3 shows the EE versus the fixed power consumption of the system, denoted by $P_f$. As expected, the EE decreases with increasing $P_f$ across all considered schemes. For any given value of $P_f$, the proposed scheme closely approaches the performance of the exhaustive search method and consistently outperforms all other benchmark schemes, thereby reaffirming its effectiveness.

Figure 4 depicts the variation of EE with respect to $D_x$,
under the conditions that $D_y$ is fixed and $L=D_x$. The EE achieved by both the proposed scheme and the exhaustive search remains nearly constant as $D_x$ varies. In contrast, the EE of the other benchmark schemes exhibits a declining trend with increasing $D_x$. This again shows the effectiveness of the proposed scheme. 
 

}


\begin{figure}[ht!]
\centering
\includegraphics[width=0.8\linewidth]{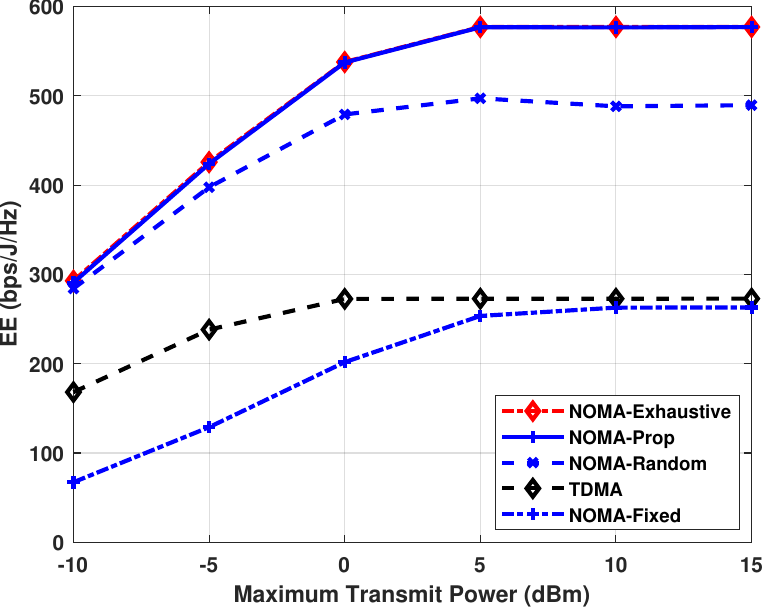}
\caption{EE versus maximum transmit power at the users.} 
\label{fig:power}
\end{figure}

\begin{figure}[ht!]
\centering
\includegraphics[width=0.8\linewidth]{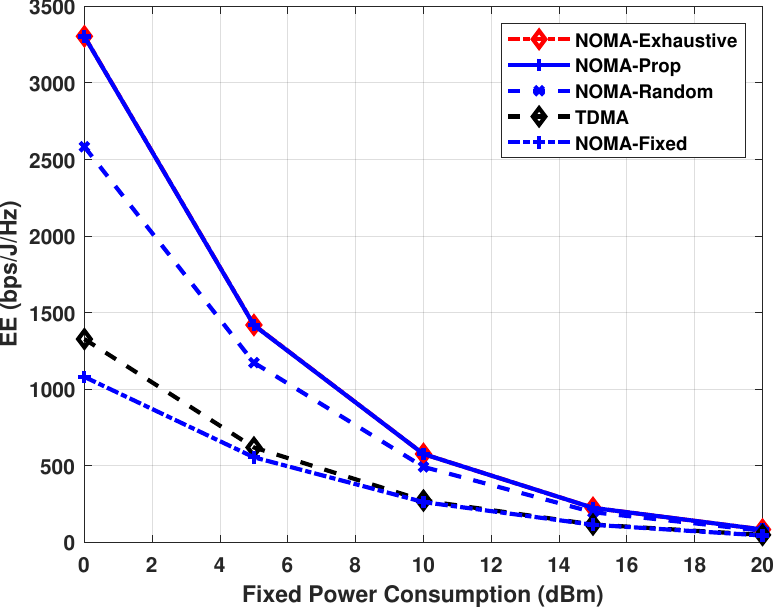}
\caption{EE versus fixed power consumption, i.e., $P_f$.} 
\label{fig:power}
\end{figure}

\begin{figure}[ht!]
\centering
\includegraphics[width=0.8\linewidth]{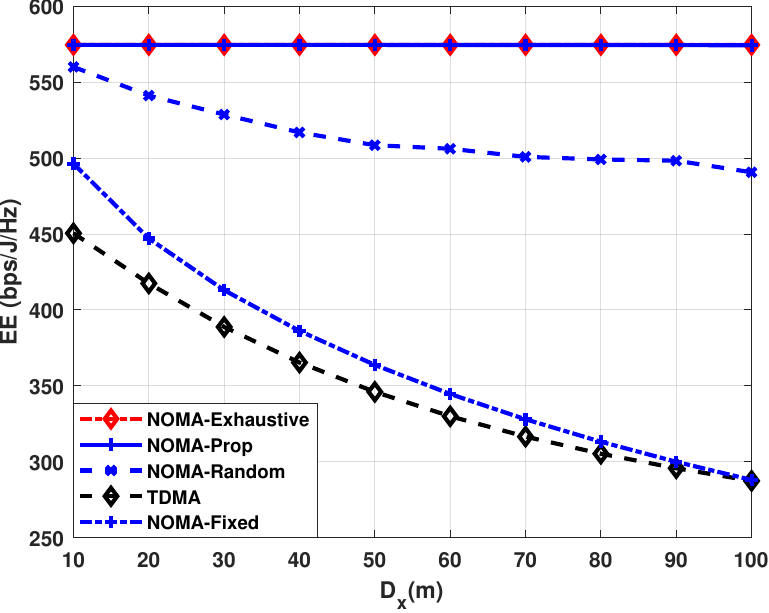}
\caption{EE versus $D_x$, with $L=D_x$ and $D_y=20$.} 
\label{fig:power}
\end{figure}

\section{Conclusion} 
\label{Sec:Conclusion}
This paper investigated the problem of maximizing EE in a NOMA-assisted uplink system utilizing a pinching-antenna configuration. The formulated optimization problem was inherently non-convex due to the coupling of decision variables. To tackle this challenge, we employed an AO framework with an appropriate initialization strategy, decomposing the original problem into two tractable subproblems: one dedicated to transmit power allocation and the other to optimizing the pinching-antenna position. We derived a low-complexity optimal solution for the power allocation subproblem and applied a PSO algorithm to solve the antenna positioning subproblem. Simulation results demonstrated that the proposed approach achieved near-optimal performance and outperformed benchmark schemes, including conventional-antenna configurations, TDMA, and random pinching-antenna initialization.


\bibliographystyle{IEEEtran}
\bibliography{biblio}

\begin{thebibliography}{10}
\providecommand{\url}[1]{#1}
\csname url@samestyle\endcsname
\providecommand{\newblock}{\relax}
\providecommand{\bibinfo}[2]{#2}
\providecommand{\BIBentrySTDinterwordspacing}{\spaceskip=0pt\relax}
\providecommand{\BIBentryALTinterwordstretchfactor}{4}
\providecommand{\BIBentryALTinterwordspacing}{\spaceskip=\fontdimen2\font plus
\BIBentryALTinterwordstretchfactor\fontdimen3\font minus \fontdimen4\font\relax}
\providecommand{\BIBforeignlanguage}[2]{{%
\expandafter\ifx\csname l@#1\endcsname\relax
\typeout{** WARNING: IEEEtran.bst: No hyphenation pattern has been}%
\typeout{** loaded for the language `#1'. Using the pattern for}%
\typeout{** the default language instead.}%
\else
\language=\csname l@#1\endcsname
\fi
#2}}
\providecommand{\BIBdecl}{\relax}
\BIBdecl

\bibitem{Sun_TVT18}
S.~Sun \emph{et~al.}, ``Propagation models and performance evaluation for {5G} millimeter-wave bands,'' \emph{IEEE Trans. Veh. Technol.}, vol.~67, no.~9, pp. 8422--8439, Sep. 2018.

\bibitem{Hao_Network22}
W.~Hao \emph{et~al.}, ``Ultra wideband thz {IRS} communications: Applications, challenges, key techniques, and research opportunities,'' \emph{IEEE Network}, vol.~36, no.~6, pp. 214--220, November/December 2022.

\bibitem{Atsushi_22}
A.~Fukuda \emph{et~al.}, ``Pinching antenna using a dielectric waveguide as an antenna,'' \emph{Technical Journal}, vol.~23, no.~3, pp. 5--12, Jan. 2022.

\bibitem{ding2024}
Z.~Ding \emph{et~al.}, ``Flexible-antenna systems: A pinching-antenna perspective,'' \emph{IEEE Trans. Commun}, pp. 1--1, 2025.

\bibitem{yang2025}
\BIBentryALTinterwordspacing
Z.~Yang \emph{et~al.}, ``Pinching antennas: Principles, applications and challenges,'' 2025. [Online]. Available: \url{https://arxiv.org/abs/2501.10753}
\BIBentrySTDinterwordspacing

\bibitem{liu2025pinching}
\BIBentryALTinterwordspacing
Y.~Liu \emph{et~al.}, ``Pinching-antenna systems ({PASS}): Architecture designs, opportunities, and outlook,'' 2025. [Online]. Available: \url{https://arxiv.org/abs/2501.18409}
\BIBentrySTDinterwordspacing

\bibitem{tyrovolas2025}
\BIBentryALTinterwordspacing
D.~Tyrovolas \emph{et~al.}, ``Performance analysis of pinching-antenna systems,'' 2025. [Online]. Available: \url{https://arxiv.org/abs/2502.06701}
\BIBentrySTDinterwordspacing

\bibitem{Xu_WCL25}
Y.~Xu \emph{et~al.}, ``Rate maximization for downlink pinching-antenna systems,'' \emph{IEEE Wirel. Commun. Lett.}, pp. 1--1, 2025.

\bibitem{wang2024}
K.~Wang \emph{et~al.}, ``Antenna activation for {NOMA} assisted pinching-antenna systems,'' \emph{IEEE Wirel. Commun. Lett.}, pp. 1--1, 2025.

\bibitem{xie2025}
\BIBentryALTinterwordspacing
X.~Xie \emph{et~al.}, ``A low-complexity placement design of pinching-antenna systems,'' 2025. [Online]. Available: \url{https://arxiv.org/abs/2502.14250}
\BIBentrySTDinterwordspacing

\bibitem{Zeng_COMML25}
\BIBentryALTinterwordspacing
M.~Zeng \emph{et~al.}, ``Sum rate maximization for {NOMA}-assisted uplink pinching-antenna systems,'' 2025. [Online]. Available: \url{https://arxiv.org/abs/2505.00549}
\BIBentrySTDinterwordspacing

\bibitem{fu2025}
\BIBentryALTinterwordspacing
Y.~Fu \emph{et~al.}, ``Power minimization for noma-assisted pinching antenna systems with multiple waveguides,'' 2025. [Online]. Available: \url{https://arxiv.org/abs/2503.20336}
\BIBentrySTDinterwordspacing

\bibitem{Zhang_TVT17}
Y.~Zhang \emph{et~al.}, ``Energy-efficient transmission design in non-orthogonal multiple access,'' \emph{IEEE Trans. Veh. Technol.}, vol.~66, no.~3, pp. 2852--2857, Mar. 2017.

\bibitem{Zeng_Access18}
M.~Zeng \emph{et~al.}, ``Energy-efficient power allocation for {MIMO-NOMA} with multiple users in a cluster,'' \emph{IEEE Access}, vol.~6, pp. 5170--5181, 2018.

\bibitem{Zeng_COMML21}
M.~Zeng, X.~Li, G.~Li, W.~Hao, and O.~A. Dobre, ``Sum rate maximization for {IRS}-assisted uplink {NOMA},'' \emph{IEEE Commun. Lett.}, vol.~25, no.~1, pp. 234--238, 2021.

\bibitem{Dinkelbach}
W.~Dinkelbach, ``On nonlinear fractional programming,'' \emph{Management Science}, vol.~13, no.~7, pp. 492--498, 1967.

\bibitem{Anuar_ISCAS19}
A.~Dorzhigulov and A.~P. James, ``Generalized bell-shaped membership function generation circuit for memristive neural networks,'' in \emph{Proc. IEEE ISCAS}, 2019, pp. 1--5.

\end{thebibliography}

\balance

\end{document}